\documentclass[grl]{agutex2015}

 \usepackage{graphicx}
\authorrunninghead{BEHAR ET AL.}

\titlerunninghead{MASS LOADING AT COMET 67P}

\authoraddr{Corresponding author: Etienne Behar, Swedish Institute of Space Physics, Kiruna, Sweden. (etienne.behar@irf.se)}

\begin{document}

%
%

\title{Mass loading at 67P/Churyumov-Gerasimenko: a case study.}
%
%

%
%



\authors{E. Behar\altaffilmark{1},
H. Nilsson\altaffilmark{1},
G. Stenberg Wieser\altaffilmark{1},
Z. Nemeth\altaffilmark{2},
T.W. Broiles \altaffilmark{3},
I. Richter\altaffilmark{4}}

\altaffiltext{1}{Swedish Institute of Space Physics, Box 812, 981 28 Kiruna, Sweden.}
\altaffiltext{2}{Wigner Research Centre for Physics, 1121 Konkoly Thege Street 29-33, Budapest, Hungary.}
\altaffiltext{3}{Space Science and Engineering Division, Southwest Research Institute (SwRI), 6220 Culebra Rd San Antonio, TX 78238.}
\altaffiltext{4}{Technicsche Universit\"{a}t Braunschweig, Institute for Geophysics and Extraterrestrial Physics, Mendelssohnstra\ss e 3, D-38106 Braunschweig, Germany.}

%
%


\keypoints{\item Prime role of the SW electric field in the cometary ion dynamics, through mass loading, at 2.88AU
\item The cometary ion flow direction has a main anti-sunward component
\item We find an indication for an anti-sunward polarisation electric field developing in the coma}


%
%


\begin{abstract}
We study the dynamics of the interaction between the solar wind ions and a partially ionized atmosphere around a comet, at a distance of 2.88 AU from the sun during a period of low nucleus activity. Comparing particle data and magnetic field data for a case study, we highlight the prime role of the solar wind electric field in the cometary ion dynamics. Cometary ion and solar wind proton flow directions evolve in a correlated manner, as expected from the theory of mass loading. We find that the main component of the accelerated cometary ion flow direction is along the anti-sunward direction, and not along the convective electric field direction. This is interpreted as the effect of an anti-sunward polarisation electric field adding up to the solar wind convective electric field.
\end{abstract}

%
%

%

\begin{article}

%
%

\section{Introduction}
The phenomenon of mass loading is common in space plasmas. Newly charged material added to the fast solar wind flow is accelerated by the Lorentz force. The newly added material gains energy and momentum from the solar wind. Solar wind ions experience an equal but opposite net force, thus balancing the total momentum of the system. The thin atmosphere permeated by the solar wind around a comet is one of the most evident cases where mass loading is expected to control the dynamics of the plasma environment (\citet{szego2000ssr}, section 4.1). {\it In situ} investigations of the solar wind interaction with a cometary plasma were made possible by different missions prior to Rosetta: ICE at P/Giacobini--Zinner in 1985, Giotto, Vega-1 and -2, Suisei, Sakigake at P/Halley in 1986, and Giotto at P/Grigg--Skjellerup in 1992, are some examples. However, all those measurements were performed during single flybys, at about 1 A.U. away from the Sun, and bow shocks were observed at each flyby, indicating a high nucleus activity ({\it cf.} \citet{neugebauer1990rg} and \citet{coates1997asr}).

The Rosetta mission \citep{glassmeier2007ssr} has provided a unique opportunity to continuously observe mass loading in a cometary environment over longer time scales and during varying nucleus activity. The Rosetta spacecraft reached comet 67P/Churyumov--Gerasimenko (67P/CG) in early August 2014. First results from the plasma measurements made at 67P/CG describe how the cometary environment evolves from a thin coma where only low fluxes of low energy ions are observed, to the point when the effect on the solar wind flow becomes significant \citep{nilsson2015aa}, \citep{nilsson2015science}, \citep{goldstein2015grl}. During these early observations no plasma boundaries had yet formed between the solar wind and the comet atmosphere, {\it i.e.} there was no bow shock or ionopause. The scale size of the interaction observed at comet 67P/CG was small, initial observations of water ions were made only when the spacecraft got closer to the nucleus than 100 km distance. 
The solar wind -atmosphere interaction at a low activity comet may thus have similarities to artificial comets formed through barium and lithium ion releases from the AMPTE spacecraft \citep{haerendel1986nature, rodgers1986nature, coates1986jgr, coates2015jpcs}.

For a low activity comet the solar wind is undisturbed before permeating the coma, and no other acceleration process has to be taken into account to study the solar wind - atmosphere interaction. We expect to observe the simplest mass loading phenomenon: newly charged mass is simply added to the undisturbed solar wind flow. This interaction has been previously addressed by \citet{broiles2015aa}, using data from another particle instrument within the Rosetta Plasma Consortium (RPC), RPC-IES \citep{burch2007ssr}. They reported that the solar wind near the comet was deflected by a Lorenz Force opposite to that experienced by cometary pickup ions. They also found that this deflection was not well ordered by the spacecraft position relative to the comet, and was well correlated with large changes in the observed magnetic field.

We present a case study that provides new details about the dynamics of this interaction between the solar wind and the coma, based on ion and magnetic field data from the 28th of November 2014.


\section{Instrument description}

The Ion Composition Analyzer, part of the Rosetta Plasma Consortium (RPC-ICA), is an ion spectrometer aimed to study the interaction between the solar wind and positive cometary ions at comet 67P/CG \citep{nilsson2007ssr}. The instrument resolves energy and mass per charge of the incoming ions. The energy spans from 10 eV up to 40 keV. The instrument field of view is $360^{\circ} \times 90^{\circ}$ ($azimuth\ \times\ elevation$, illustrated in Figure \ref{fov}), with a resolution of $22.5^{\circ} \times 5.0^{\circ}$. For this data set, a full angular scan was produced every 192 s. The elevation angle of the incoming positive ions is determined by an electrostatic acceptance angle filter at the entrance of the instrument, and the azimuth angle is measured by the means of 16 anodes, part of the detection system marking the end of an ion path in the instrument. Thus the two angles are subject to different constraints, limits, and resolutions.

The magnetometer (RPC-MAG, \cite{glassmeier2007ssr-MAG}) measures the three components of the magnetic field vector in the range from DC up to 10 Hz. The measurement range is $\pm$ 16384 nT with a resolution of 31 pT. RPC-MAG is mounted on a 1.5 m long boom in order to minimise the impact of the spacecraft generated disturbance fields. The magnetometer is affected by a systematic bias field from the spacecraft, which cannot  be fully characterized as the spacecraft fields are changing related to the operation status. In this work, we take this into account by considering and propagating a $\pm$ 3 nT uncertainty on each magnetic field component in the spacecraft reference frame.


\section{The case}

In order to diagnose the dynamics of the interaction between the solar wind and the coma, we considered data sets with a clear and constant water ion signal, simultaneously with a clear solar wind signal. The chosen data set turned out to be the clearest in terms of dynamics, mainly because of large variations in the upstream magnetic field direction.

The following case study is based on particle and magnetic field data collected at 2.88 AU on the 28th of November 2014. The spacecraft was then flying a terminator orbit 30 km away from the center of the nucleus, with a mean speed over the day of 0.15 m/s relative to the comet. This very low speed allows us to neglect any aberration angle concerning the cometary ions.

Moreover, an aberration of 1.01$^{\circ}$ is obtained for a 400 km/s fast solar wind. Our angular resolution is larger than this angle, so we neglect this aberration angle as well.

To study the dynamics of the interaction, we focus on two species: solar wind protons $H^+$ and cometary water ions $H_2O^+$, the most abundant species in the solar wind and the ionized atmosphere respectively \citep{nilsson2015science}. A large flux of accelerated water ions is observed during the chosen day, and we consider the energy range [70 eV, 330 eV], with a peak value at 100 eV. The upper bound is set higher than the most energetic water ions observed. The lower bound isolates these accelerated water ions from the cold water ion population reported by \citet{nilsson2015aa}: this population is affected by the spacecraft potential in terms of direction, and therefore is not physically relevant for our study. Protons are observed in the energy range [350 eV, 1200 eV]. 

   The average magnetic field magnitude is 14 nT $\pm$ 5.2 nT over the 14 hours of the data set. The speed of the observed protons is stable, with a value of 380 km/s for a peak at 750 eV and a standard deviation $\sigma$ = 20 km/s. 

 As a first approximation, we assume that the solar wind electric field is given by ${\bf E} = -{\bf v}_{H^+} \times {\bf B}$. Integrating the movement of a test particle with a velocity $v_{test}$ ,  only subject to the Lorentz force ${\bf F} = q({\bf E} + {\bf v_{test}} \times {\bf B}) $, we estimate that the most energetic cometary ions observed ($\sim$ 300 eV) were accelerated during less than 3 s, over a distance of about 40 km. With this approximation, the minimum gyro-radius is 2600 km, which means that the $H_2O^+$ we observe are on the very early phase of the gyro-motion, {\it i.e.} flowing along the local electric field. For this reason, we expect that the observed accelerated water ion flow gives us the direction of the local electric field. These numbers are summarized in Table \ref{gyro}.
 
 To complete the description of the environment during the measurement, we estimate the density profile of the coma using the Haser model (\citet{haser1957}). In this model, the neutral density falls off following $\sim 1/r^2$, with $r$ the distance to the nucleus. At 100 km away from the nucleus, the densities would be an order of magnitude lower than the one met along the 30 km terminator orbite.

\begin{table}
\caption{Conditions and estimations of $H_2O^+$ gyro-radius \label{gyro}}
\centering
\begin{tabular}{r | c | c}

   & Data & Undisturbed  \\
   && solar wind  \\
\hline
\hline
  $B$ & [12, 20] nT & 1 nT \\
  $v_{H^+}$ & 380 km/s & 400\ km/s \\
  $ E_{estim}$ & [4.0,\ 4.6] V/km & 3.7 V/km \\
\hline
  $R_{gyro}$ & [2600, 5900] km & -
\end{tabular}

\end{table}


\section{Method}

To study the dynamics of the interaction between the solar wind and the comet atmosphere, we aim to express the flow directions of $H^+$ and $H_2O^+$ in the body-Centered Solar EQuatorial (CSEQ) frame. To achieve that, we first collect the observed counts for the two ion populations, which are well separated in both energy and mass. We then compute full angular distributions for each species every 192 s. 

\begin{figure*}
   \begin{center}
      \includegraphics[width=\textwidth]{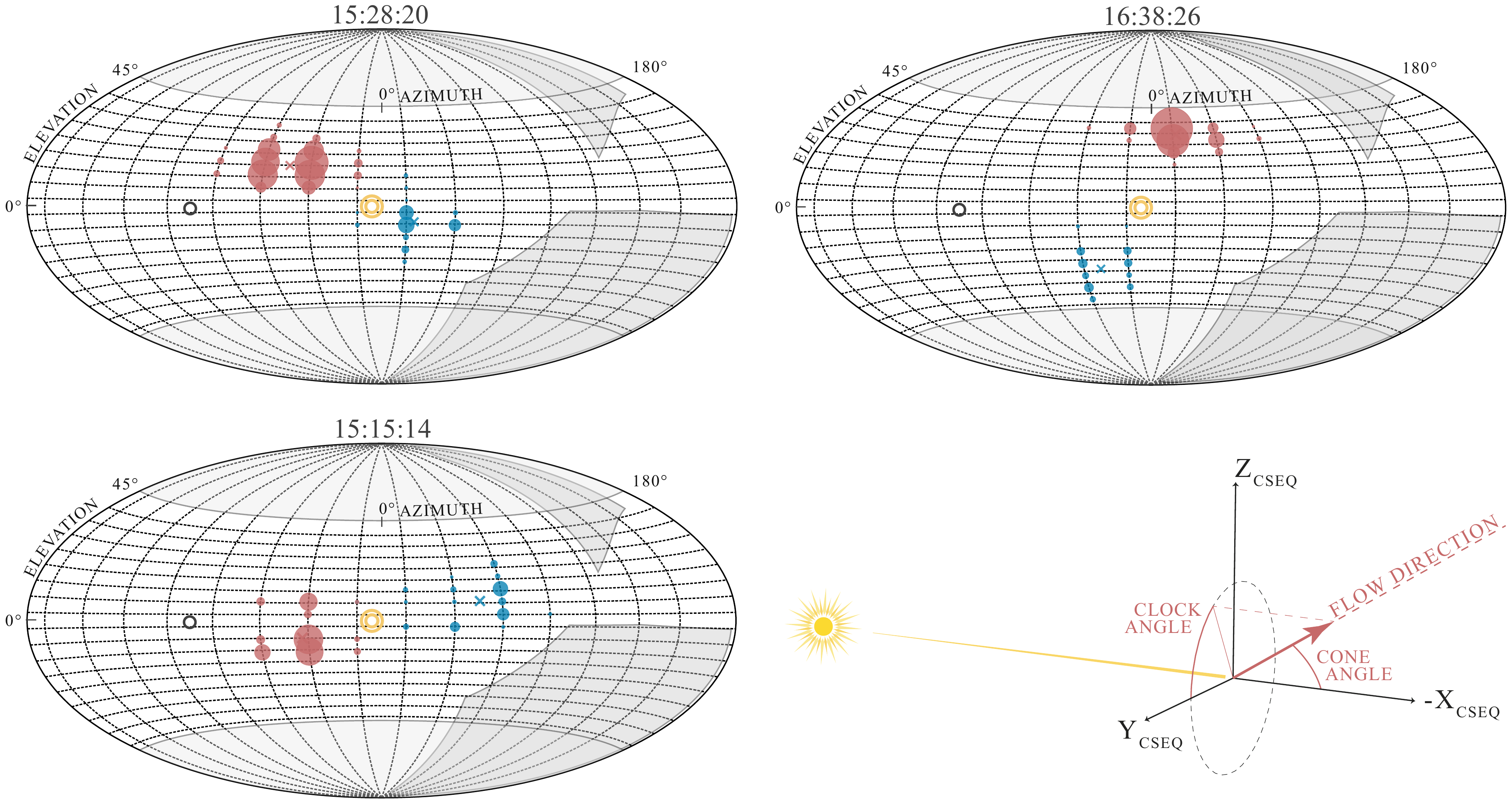}
      \caption{Three examples of $H_2O^+$ and $H^+$ flows pictured in the instrument FOV at precise times during the 28th of November. Depending on the time, the two flows are coming from completely different directions (the spacecraft attitude is unchanged during this period), but always with a $180 ^{\circ}$ difference in clock angle as defined in the lower right illustration. It represents the spherical coordinate system used for this study. All vectors are normalized, ${\bf v}_{norm.} = \frac{\bf v}{|{\bf v}|}$ \label{fov}}
   \end{center}
\end{figure*}

To clearly visualize the two flows and their dynamics, we produce a sequence of pictures as seen with the instrument Field Of View (FOV). Three of the pictures composing the sequence are given in Figure \ref{fov}; the full sequence can be seen at http://irf.se/$\sim$etienne/mediaFOV.html . To that extent, the instrument is used as a camera. The produced sequence also helps us assess that the restricted field-of-view of the instrument does not impact our results.

We calculate the direction of the bulk velocity for both species, then express the two flow directions (unit vectors) in the CSEQ reference frame. The $x_{CSEQ}$ axis is along the sun-comet line, pointing to the sun. The $z_{CSEQ}$ axis is parallel to and oriented by the Sun's north pole direction, orthogonal to the +X axis. The $y_{CSEQ}$ completes the right-handed reference frame. We compute the angle of each flow direction from the sun-comet line, and refer to this angle as the cone angle. We complement it with the clock angle expressed in the $(y_{CSEQ}, \ z_{CSEQ})$ plane. It is the angle of the projected flow direction in this plane, from the $y_{CSEQ}$ axis and positive towards $z_{CSEQ}$, as shown in Figure \ref{fov}, lower right illustration. Both angles form a spherical coordinate system, and all vectors are normalized in this study, ${\bf v}_{norm.} = \frac{\bf v}{|{\bf v}|}$. The magnetic field is also expressed in this coordinate system.\\


\section{Results and discussion}

   If mass loading is the only mechanism in the interaction between the solar wind and the coma, we expect to see cometary ions accelerated along the local electric field direction and solar wind protons deflected with an opposite clock angle, as a result of momentum conservation. The dynamics takes place in the plane that includes both flow directions and the comet-sun line. As the direction of the solar wind electric field varies, both flows are expected to follow the field rotation, $180 ^{\circ}$ away from each other in terms of clock angle. We thus verify with only one coordinate if the two ion populations flow in the same plane or not. The cone angle quantifies the anti-sunward component in this plane. It gives the amount of solar wind proton deflection, and the anti-sunward component of the cometary ion velocity.

   \subsection{Clock angle, momentum conservation}

   In Figure \ref{comp}, clock angles for $H^+$ (red dots),  $H_2O^+$ (blue dots)  and ${\bf B}$ (black line) are given in the top panel.  All three clock angles show variations over a span of about  $200 ^{\circ}$. Despite these large variations, the difference between proton and cometary ion flow directions remain around  $180 ^{\circ}$, and is given in the center panel. 
   The mean value of this angular difference is $187 ^{\circ}$ and its standard deviation is $18 ^{\circ}$. The proton flow clock angle is more correlated to the magnetic field clock angle than the cometary ion flow is. The variations thus mainly originate from the cometary ion flow direction, which is most of the time less regular, or beam-like ({\it cf.} Figure \ref{fov}, lower left FOV picture: there are two peak values instead of one). An energy dependence in the water ion flow direction could explain these variations, and will be the topic of a more advanced study focused on the cometary ion population.

   The two observed flows have very different directions. Thus the two populations, $H^+$ and $H_2O^+$, have been flowing along different paths in the coma. Observing this $180 ^{\circ}$ difference ensures us that the magnetic and electric fields have respectively very similar directions along these two different paths. As the upstream electric field direction varies, the two paths  also move inside the coma, but the angular difference remains the same. This also confirms that the flow directions of both populations remain in the same plane, containing the comet-sun line. 
   
The direction of the deflection doesn't seem to be influenced by the position of the nucleus, the flow may be deflected away just as well as towards the nucleus. There is no indication of the ion population flowing around the nucleus. This is because the size of the the nucleus is much smaller than the gyroradii of solar wind ions. To be able to flow around an obstacle, the particles of the flow must have sufficient time  and space to interact several times inside a boundary layer. The interactions inside the boundary layer mediate the effect which "pushes away" the flow in a conventional situation. There is no such layer in the case of the new born cometary magnetosphere. The flow only experiences electromagnetic forces due to the mass loading, it feels the much larger inner coma that way, but it simply cannot feel the nucleus. Solar wind ions are bombarding the surface of the nucleus instead of flowing around it. In this regard this early stage of the interaction is very different from later stages.
 
   The dynamics is clearly driven by the solar wind electric and magnetic fields. The nearly constant difference in clock angle between $H_2O^+$ and $H^+$ and the correlation with magnetic field clock angles emphasize the role of mass loading as the main mechanism controlling the dynamics of this plasma environment.

   \subsection{Cone angle}

   The cone angles of $H^+$, $H_2O^+$ and ${\bf B}$ are given in the third panel of Figure \ref{comp}. An anti-correlation between the $H_2O^+$ flow direction and the magnetic field can be seen: when the magnetic field cone angle decreases (i.e. the magnetic field is more parallel to the sun-comet line), the $H_2O^+$ cone angle increases, and vice-versa.
    
      Concerning the proton flow, the cone angle seems to be correlated with the magnetic field cone angle in the first half of the data set. Around 9:40, a sudden and large decrease is observed in both angles. This correlation has an immediate interpretation. $B_\perp$ is the component of the magnetic field orthogonal to the proton flow. The larger $B_{\perp}$ is (for the same $B$ amplitude), the larger ${\bf E} = -{\bf v}_{H^+} \times {\bf B} $ is. The acceleration of new born ions is then larger and as a direct consequence the deflection of the protons is also larger. The correlation is not clearly observed after 12:00. 
       
\begin{figure}
   \begin{center}
      \includegraphics[width=.5\textwidth]{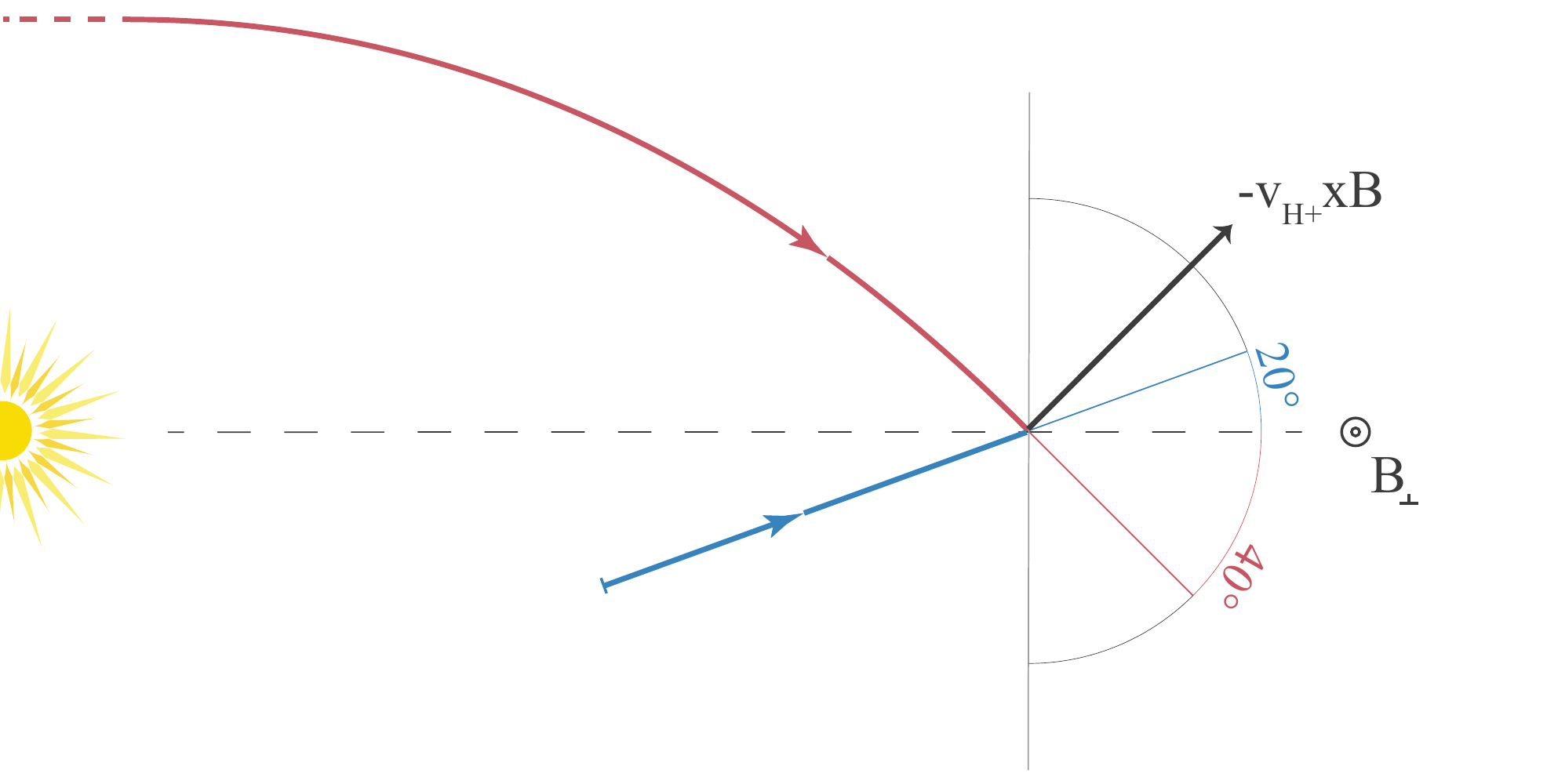}
      \caption{Illustration of the observation: the solar wind protons (red line) are gradually deflected by the coma, reaching a $40 ^{\circ}$ angle from the comet-sun line in average, measured in the terminator plane. The main component of the cometary ion flow is along the comet-sun line. The dynamics takes place in a plane, containing the comet-sun line and rotating around it. \label{result} }
   \end{center}
\end{figure}

     On average, $H_2O^+$ flow direction is $20 ^{\circ}$ away from the sun-comet line, and a $40 ^{\circ}$ solar wind proton deflection is observed, as illustrated in Figure \ref{result}.\\

      \subsection{Electric field}
      
     The generalized Ohm's Law can be written as following:
      
      \[ {\bf E} = {\bf E}_{motional} + {\bf E}_{other}\] 
      
      As previously introduced, we approximate the direction of the motional field by ${\bf E}_{motional} = -{\bf v}_{H^+} \times {\bf B}$, computing together particle data and magnetic field data. For two earlier cases (data from 2014-09-21 --{\it cf.} \citet{nilsson2015science}-- and 2014-11-16), the $H_2O^+$ ions are flowing along $-{\bf v}_{H^+} \times {\bf B}$ . ${\bf E} = {\bf E}_{motional} = -{\bf v}_{H^+} \times {\bf B}$ was then a good description of the local electric field. These data sets correspond to a lower activity and larger distances to the nucleus, resulting in a lighter mass loading of the solar wind. In the extreme case where only few cometary ions are picked up, the solar wind is almost undisturbed by the interaction, and those few pick up ions are first accelerated along the convective electric field given by ${\bf E} = -{\bf v}_{H^+} \times {\bf B}$. 
      
\begin{figure*}
   \begin{center}
      \includegraphics[width=\textwidth]{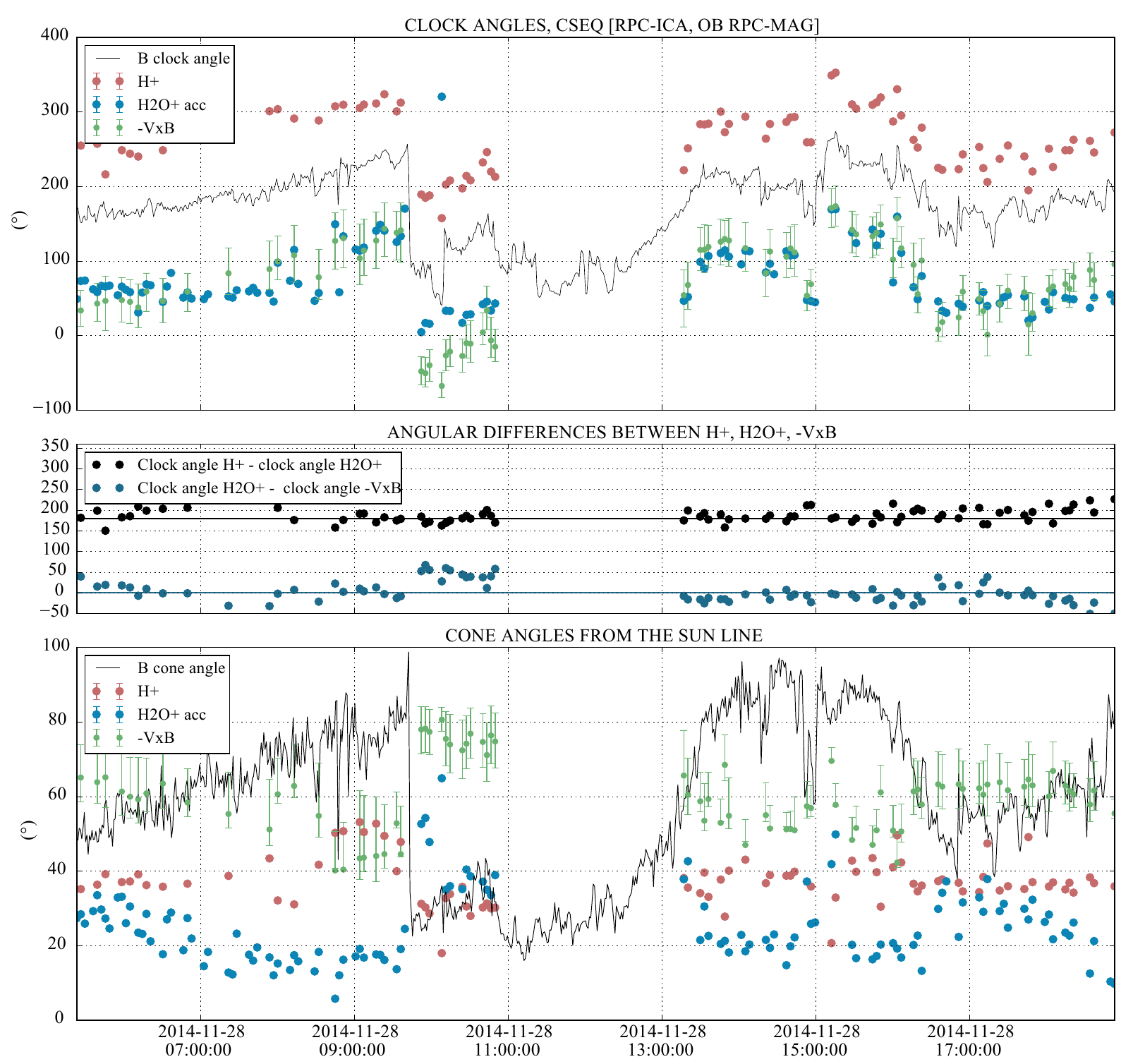}
      \caption{Clock (top panel) and cone (bottom panel) angle time series, for the $H^+$(red) and $H_2O^+$ (blue) flow directions, and for the magnetic field direction (solid black line). The difference between the two flow clock angles is given (center panel), with the two solid lines indicating $y=180 ^{\circ}$ and $y=0 ^{\circ}$. Finally, the $-{\bf v}_{H^+} \times {\bf B}$ direction is given in green, and its angular difference with cometary ion flow is given in the center panel in terms of clock angle. Between 11:00 and 13:00, no particle data are available. \label{comp}}
   \end{center}
\end{figure*}

       The case studied here corresponds to a more significant mass loading of the solar wind and reveals a new configuration of the electric and magnetic fields inside the coma. The result for the comparison between the $-{\bf v}_{H^+} \times {\bf B} $ direction and the $H_2O^+$ flow direction is given in Figure \ref{comp}, using the same spherical coordinate system previously introduced. In terms of clock angle (first panel), the two series are overlapping nicely. Their angular difference, given in the second panel, is centered around the solid black line at $0^{\circ}$. We note that between 09:45 and 11:00, after the sudden decrease in the magnetic field cone angle, this difference is centered around $ \sim 45 ^{\circ}$. This may indicate that a further specific study of this quasi-parallel regime is needed.
     
     The cone angles give us a more complex picture of the dynamics. $-{\bf v}_{H^+} \times {\bf B} $ and $H_2O^+$ flow cone angles seem to correlate with each other, but with an angular difference of about $30 ^{\circ}$. So it appears that with the activity increasing, ${\bf E} = -{\bf v}_{H^+} \times {\bf B}$ doesn't hold anymore.  Two possibilities can explain the discrepancy:
      \begin{itemize}
      \item In the description of the motional electric field ${\bf E}_{motional} = -{\bf v} \times {\bf B}$ , ${\bf v_{H^+}}$ is not a good proxy anymore. Electron and proton flows do not have the same direction anymore.
      \item ${\bf E}_{other}$ became comparable to ${\bf E}_{motional}$ .
      \end{itemize}
      
      Both possibilities can be directly put in relation with the increasing cometary ion density, and it seems reasonable to say that both are playing a role. The magnetic field is enhanced (with an average value of 14 nT, compared to the $\sim 1$ nT expected in the undisturbed solar wind at this heliocentric distance), while the solar wind protons are not significantly slowed down (the observed proton flow speed is 380 km/s). This implies that the electron fluid is significantly slowed down. There is no obvious reason why this should lead to an anti-sunward electric field. Therefore, we don't explore this first possibility, and confine the discussion to the possibility of another electric field developing in the coma when densities are higher, resulting in an anti-sunward acceleration of picked up cometary ions.\\

    The cometary atmosphere provides a constant and distributed source of ions and electrons. The electrons are picked up and swept downstream (${\bf E} \times {\bf B}$ drift), whereas the new-born ions are moving along the local electric field direction. Charge separation occurs because of the scale size of the interaction region being much smaller than the gyro radius of a new born water ions. Whatever electric fields arise from this, must in the end have the net effect of reducing this separation. The new net polarisation field is in the plane $ ({\bf v}_{H_2O^+}, {\bf v}={\bf E} \times {\bf B})$ and therefore cannot have any influence on the clock angle of the $H_2O^+$  and $H^+$ ion flow directions. It has however a large influence on their cone angle, as seen in the observation presented in the previous sub-section.
    
    If we now consider the solar wind, we expect that ions and electrons will also react differently to the obstacle on some scale. As previously mentioned, observations show no significant slowing down of the solar wind while we see a significant increase in the magnetic field as compared to the undisturbed solar wind. That implies decoupling between electrons and protons, which could lead to a Hall current or even charge separation.
    
    To get the full picture of this interaction and its dynamics, one would need to add the two components discussed here-above. The situation becomes very complex, and we believe that numerical simulations are needed to describe the interaction. We observe a mainly anti-sunward cometary ion flow, implying a mainly anti-sunward $E_{other}$. This is consistant with the mechanism discussed in the first paragraph. 
    
     \cite{haerendel1986nature} depict such an anti-sunward electric field in the context of the AMPTE artificial comet, introducing it with the same argument of charge separation. This work was based on magnetic field data and particle measurements made by a similar 3D plasma instrument ({\it cf.} \citet{rodgers1986nature} and \citet{coates1986jgr}). This topic is also widely reviewed by \citet{szego2000ssr}. 
      However, the AMPTE releases concerned transient phenomena, [3] whereas Rosetta investigate the evolution of a continuous and time varying source of neutral particles.


\section{Conclusion}

   Despite a limited angular resolution and uncertainties on the magnetic field components, the obvious correlations and anti-correlations presented in this study provide a detailed description of the interaction between the solar wind and the cometary atmosphere, during a period of low nucleus activity. 
   The plasma environment dynamics pictured in this work does not depend on the nucleus position: when the solar wind magnetic field direction changes, the two flow directions evolve accordingly, regardless of the nucleus position, {\it i.e.} the ions don't flow around the nucleus. This is in agreement with earlier results, {\it cf.} \citet{broiles2015aa}. Both ion populations -- from the solar wind and from the coma -- flow with clock angles $180 ^{\circ}$ apart from each other: the dynamics of the interaction takes place in a single plane, that contains and rotates around the comet-sun line.

   Measured at 30 km away from the nucleus and at a distance of 2.88 AU from the Sun, the solar wind protons reach deflections higher than $50 ^{\circ}$. Even in the absence of plasma boundaries, the solar wind is already substantially disturbed, and the magnetic field is not coupled to the solar wind ions anymore, contrary to what is observed at greater distances to the Sun.

The comet ions are accelerated with a dominating anti-sunward component. We suggest that the difference of motion between the cometary ions and electrons, together with the limited scale size of the coma, smaller than the cometary ion gyro-radius, result in a mainly anti-sunward new contribution to the electric field.


%
%
%
%
%
%
%

\begin{acknowledgments}
The work on RPC-ICA, as well as this PhD project, is founded by the Swedish National Space Board.
The work on RPC-MAG was financially supported by the German Ministerium f\"ur Wirtschaft und Energie and the Deutsches Zentrum f\"ur Luft- und Raumfahrt under contract 50QP 1401 .
Without the tremendous work of the Rosetta Science Ground Segment (RSGS), Rosetta Mission Operation Control (RMOC) and all instrument team planners, this study would be impossible.
Sharing data within RPC is made possible by the web-based interface AMDA, developed and made available for RPC use by Centre de Donn\'ees de la Physique des Plasmas (CDPP). This efficient interface has been of a great use for this work.
The SPICE toolkit from NASA's Navigation and Ancillary Information Facility (NAIF) is heavily used in this study, and made accessible through python by the SpiceyPy wrapper.
\end{acknowledgments}

\end{article}
%
%
%
%
%
%
%
%


\end{document}